# Application of a Regional Model to Astronomical Site Testing in Western Antarctica

**Mark Falvey[1] and Patricio M. Rojo[2]**

**Abstract.** The quality of ground based astronomical observations are significantly affected by local atmospheric conditions, and the search for the best sites has led to the construction of observatories at increasingly remote locations, including recent initiatives on the high plateaus of East Antarctica where the calm, dry and cloud free conditions during winter are recognized as amongst the best in the world. Site selection is an important phase of any observatory development project, and candidate sites must be tested in the field with specialized equipment, a process both time consuming and costly. A potential means of screening site locations before embarking on field testing is through the use of regional climate models (RCM). In this study we describe the application of the Polar version of the Weather Research and Forecast (WRF) model to the preliminary site suitability assessment of a hitherto unstudied region in West Antarctica.

Numerical simulations with WRF were carried out for the winter (MJJA) of 2011 at 3 km and 1 km spatial resolution over a region centered on the Ellsworth mountain range. Comparison with observations of surface wind speed and direction, temperature and specific humidity at nine automatic weather stations indicate that the model performed well in capturing the mean values and time variability of these variables. Credible features revealed by the model include zones of high winds over the southernmost part of the Ellsworth Mountains, a deep thermal inversion over the Ronne-Fincher Ice Shelf and strong west to east moisture gradient across the entire study area. Comparison of simulated cloud fraction with a CALIPSO spacebourne Lidar climatology indicates that the model may underestimate cloud occurrence, a problem that has been noted in previous studies. A simple scoring system was applied to reveal the most promising locations. The results of this study indicate that the WRF model is capable of providing useful guidance during the initial site selection stage of project development.

## 1. Introduction

All ground based astronomical observing systems must make use of signals that are to some extent modified by the earth's atmosphere, whether it be due to absorption by clouds and aerosols, blurring due to atmospheric turbulence or signal contamination from thermal or water vapor emissions. As astronomers have strived to produce images of ever more outstanding quality, the techniques used to compensate for or mitigate atmospheric noise have grown in scope and sophistication. It would not be an overstatement to say that "astro-meteorology" now exists as sub-discipline in its own right, with active areas of research in the modeling of turbulence (e.g., Abahamid at al, 2004), development of atmospheric correction technologies (e.g., Ellerbroek 1994), the development of targeted monitoring systems (e.g., Vernin and Muñó-Tuñón, 1995, Tokovinin and Kornilov, 2007) and the use of predictive models for the optimization of telescope usage in major observatories (Erasmus and Sarazin, 2002, Buckley 2015). The geographical location of astronomical observatories is also strongly determined by atmospheric constraints. Indeed, the quest to minimize atmospheric degradation has driven observatories to increasingly isolated, high altitude locations such as Hawaii's volcanoes or the mountains of Chile's Atacama Desert (Kerber et al, 2012). In recent years, this trend has gone to further extremes with the advent of observatories in perhaps the most remote and challenging location of all, Antarctica, where clear skies, a calm atmosphere and long night have prompted a surge in astronomical activities (Burton, 2010).

Most major telescopes enter into construction only after an extended site testing phase, in which the atmospheric conditions at one or more candidate sites are carefully measured (usually with highly specialized equipment) and compared (Lombardi 2009, Schöck et al 2009). The site testing stage may be both costly and time consuming. By way of example, the definitive location of the Thirty Meter Telescope (TMT) at Maunakea in Hawaii was determined only after an extensive measurement campaign at five locations in four continents throughout a five-year data collection period (Schöck et al 2009). Given the costs involved in the site evaluation phase, the test sites must be very carefully selected. The TMT project chose its candidate sites largely on the basis of prior measurements, local knowledge, revision of satellite cloud cover and a priori logistical concerns. In Antarctica, most current sites of interest are situated at existing scientific bases where prior measurements are available, although at least one site (Ridge A) has been identified indirectly on the basis of weather model and satellite products (Saunders et el 2009).

---

[1] Department of Geophysics, University of Chile
[2] Department of Astronomy, University of Chile. Corresponding Author Address: Department of Astronomy, University of Chile, Camino El Observatorio 1515, Las Condes, Santiago, Chile

A potential aid in site selection process is the application of regional climate models (RCM's). Modern RCM's provide complete, physically consistent description of the atmospheric circulation over a given area with high resolution both in the space and time domain. Such models may be of value in regions with limited observational coverage or complex topography, where available data cannot adequately capture the spatial patterns of relevant meteorological variables. So far, there has been little published work on the possibility of narrowing down potential observatory sites using regional models (recent examples include Masciadri and Lascaux (2012) or Giordano et al, 2014). However, the use of such models for site selection purposes well established in other contexts, a good example being the prospection of wind energy where it is common practice for project developers to first evaluate a region of interest using an atmospheric model before beginning targeted measurement campaigns at specific locations (e.g., Probst and Cardenas, 2010).

In this study we apply an RCM to the task of preliminary site screening in West Antarctica (WA). Until now astronomy in Antarctica has focused on the high ice domes on the eastern side of the continent. In contrast, WA has received little attention, presumably due its lower mean topography and supposedly more maritime climate compared to East Antarctica (EA). Nonetheless, there are reasons to suspect that parts of WA may offer unique advantages. For instance, the highest mountain of Antarctica, Mt Vinson (4892 m), is found within WA in the Ellsworth mountain range. Furthermore, the presence of mountain ranges in WA offers the possibility of positioning telescopes on solid rock surfaces. In 2011 the Chilean government announced the establishment of new Antarctic Science base in the Union Glacier, over 1000 km from the coast and within striking distance of the Vinson Massif and several other mountain ranges. The presence of the scientific base means that an astronomical observing site in this part of WA could be logistically feasible.

Here we apply the Polar version of Weather Research and Forecasting RCM (Hines and Bromwich, 2008) to simulate the spatial patterns of several key atmospheric variables that determine the site suitability. We focus on a roughly 1000 x 1000 km region centered near the Union Glacier. Our principal goal is to demonstrate how the WRF model simulations may provide a useful means of identifying potential observatory sites in given region. Although our focus region is West Antarctica, the methodology ought to be transferrable to other parts of the world. Our work includes the evaluation of model performance by comparing with available surface and satellite observations. To our knowledge, this is the first time that the WRF model has been applied to this part of Antarctica at high resolution, and as such the results may be of general interest to readers in working on Antarctic meteorology or related disciples such as glaciology or climatology.

The structure of this paper is as follows: In the next section we provide a brief discussion of the atmospheric factors that degrade astronomical observations and the aspects of Antarctica's climate that make it favorable for astronomy. In section 3 we describe the WRF regional modeling system. The observational data, which include surface observations and CALIPSO satellite cloud products are described in section 4. In section 5 the results of the simulations are presented and compared with available observations. Section 6 presents a simple spatial analysis scheme designed to condense model results into a single measure of site suitability and reveal the most promising sites in the study area. We close the paper in section 7 with a summary and discussion of the most important results.

## 2. Meteorological Factors and the Antarctic Advantage

The best astronomical sites in the world are chosen for a variety of factors and different projects have different needs. The ideal conditions for almost all projects are clear skies, calm stable conditions, and low humidity. During the Antarctic winter, these conditions have been shown to be met at several sites on the high ice plateaus of East Antarctica (e.g., Domes A, C, E and F; Burton, 2010, Bonner et al., 2010).

The aspect that has attracted most attention has been the promise of exceptionally good image sharpness, or "seeing", based on mostly indirect estimates derived from DIMM (Differential Image Motion Monitor, e.g., Aristidi et al., 2005a) or MASS (Multi-Aperture Scintillation Sensor, e.g. Lawrence et al. 2004) measurements. Seeing is measured as the angular blurring of an astronomical point source. The best observatory sites in Chile reach a best decile of 0.41 arc-seconds and a median of 0.64 arc-seconds (Skidmore et al 2009, Cerro Armazones). In Antarctica, unprecedented values below 0.2 arc-seconds have been inferred for the atmosphere above 30 m height at Dome C based on combined MASS (scintillation and SODAR (Sonic Detection And Ranging) measurements (Lawrence et al., 2004). In practice, however, it has been difficult to reproduce those values in-situ due to the existence of a turbulent boundary layer in the lowest 10-30 m (Agabi et al, 2006), which greatly degrades the seeing obtained for instruments located near ground level. At the South Pole for example, seeing conditions are usually much worse than major non-Antarctic observatories as the boundary layer extends to well over 100m in height (Marks et al 1999). Above the boundary layer, exceptional seeing conditions and stability are expected to be present over most of the continent (Saunders et al, 2009). Finding areas where the boundary layer is thin enough to allow telescopes to be placed above it is therefore an important aspect of astronomical site selection in Antarctica.

Furthermore, many potential sites over the Antarctic continent must be discounted due the presence of very strong winds in the boundary layer (the so called Katabatic winds) that develop due to the extreme cooling of the near surface air over the icefields. These winds are also associated with turbulence and degrade image quality. Katabatic flows tend to develop the on slopes of the major Antarctic landforms. Somewhat counter-intuitively, the high points (peaks, ridges and domes) of these landforms (where the topography is locally flat) tend to be rather calm and thus stand out as good candidates for astronomical observatories.

The Antarctic advantage for astronomy goes beyond the calm and clear atmospheric conditions. The cold atmosphere has very little capacity to retain vapor and the precipitable water vapor (PWV) is very low (Swain and Gallée, 2006). The lower pressure and thinner atmosphere at high latitudes further contribute to lower atmospheric contamination for ground-based observations. The extremely low temperatures reduce background emission noise from the instrumentation. Additionally, the long winter night provides a unique observing cadence, where the observation of a single object can continue for many days without interruption.

There are of course several major challenges associated with astronomy in Antarctica. Apart from obvious logistical difficulties, the lack of a solid rock implies that all current telescopes placed on ice surfaces that are continuously moving. A moving floor raises the construction costs and adds significant technical difficulties to achieve accurate astronomical pointing. Furthermore, the existence of a turbulent boundary layer requires telescopes to be mounted on top of tall towers to obtain the best conditions.

In their 2009 paper entitled "Where Is the Best Site on Earth? Domes A, B, C,and F, and Ridges A and B", Saunders et al provide a comprehensive overview of the relative merits of potential observatory sites in Antarctica. As can be readily inferred from the title of their article, Saunders et al's work focused on the evaluating the EA plateau. They considered a variety of atmospheric constraints which may be loosely divided into two categories. The first are large-scale factors that mainly depend on the broad characteristics of the upper atmosphere over Antarctica. These factors include the aurora, sky visibility and free air turbulence (which is to a large extent a function of upper air winds). Secondly, they considered several local scale factors that may vary considerably over relatively short distances, especially in regions of complex topography. These include boundary layer height, surface winds, integrated water vapor and air temperature. In some cases, local variability may be fairly predictable. For example, water vapor and air temperature decline with altitude and their spatial pattern depends mainly on the height of the surface topography. The other factors, in particular boundary layer winds and turbulence, and cloudiness, may exhibit more complex variability over small spatial scales. Although Saunders et al´s, work is comprehensive, the meteorological databases they employed were of low resolution and little attention was given to West Antarctica. In this paper we focus principally on the small scale factors mentioned previously (see table 1), which can be simulated at high resolution with a regional model and may be expected to exhibit strong small-scale variability within the study area described in the next section.

| Variable | Units | Related to | Observing type affected | Validation data |
|---|---|---|---|---|
| Surface wind speed | m/s | Seeing | Optical and near-infrared | Surface weather stations |
| Boundary layer height | mm | Seeing | Optical and near-infrared | None |
| Cloud fraction (time) | % | Potential observing time | All | CALIPSO cloud climatology |
| Integrated water vapor | mm | Signal refraction in radio frequencies, background emissions at infrared frequencies | Radio, near infrared | None. (Indirectly with surface humidity) |
| Surface temperature | °C | Background thermal emissions | Near infrared | Surface weather stations |

**Table 1.** Summary of the variables used to determine site suitably in this study. This list is not exhaustive, as there are many other factors (aerosols, sky brightness, amongst others) that may also play an important role. The variables chosen in this study are those that are expected to display significant small scale patterns over the study area and that are able to be simulated with a regional climate model. The column "Related to" indicates the relationship between the atmospheric variable and the specific signal degradation phenomena. The column "Observing type affected" indicates the specific type or types of astronomical observations that are affected by the variable. Finally, the column "Validation data" indicates the source of observations, if any, that may be used to validate the RCM simulations.

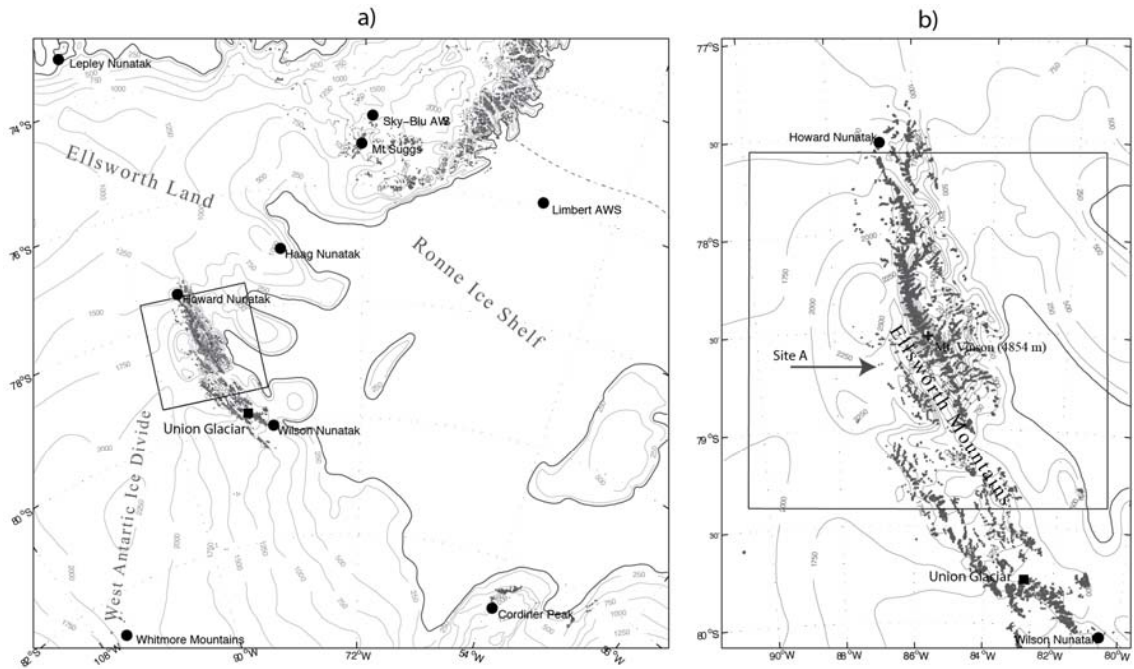

**Figure 1.** a) Topographical map of the study area. Contours show the terrain height at 250m intervals based on data from the RAMP digital elevation database. Weather stations are indicated with a black circle. The Union Glacier logistical hub is marked with a black square. Grey regions are areas of exposed rock surface taken from the SCAR Antarctic Digital Database. The area shown corresponds exactly to the 3 km resolution WRF simulation domain. The rectangle shows the location of the higher resolution 1 km simulation domain centered over the Ellsworth mountains. b) Same as a) but zooming in on the Ellsworth mountain ranges.

## 3. Study area

The study area, shown in Figure 1a, is located in the eastern sector of West Antarctica and stretches from the base of the Antarctic Peninsula (72°S) to the southern flank of the west Antarctic dome at 84°S, just 800km from the South Pole. The area is loosely comprised of three geographical regions: Ellsworth and Palmer Land to the northwest, the Western Antarctic ridge to the southwest and the Ronne Ice Shelf to the east. At the intersection of these regions are the Ellsworth Mountains (see detail in figure 1b), a small but dramatic group of mountain ranges from which rises the Vinson Massif, the highest mountain in Antarctica (4892 m). Most other peaks in the Ellsworth Mountains are much lower, but nonetheless there are many of over 2000 m in height. The Ellsworth Mountains are to a large extent ice covered, and the mountaintops often protrude only a few tens of meters from the surrounding glaciers. These features, known as Nunataks, are considered to be primary candidates for astronomy sites as they potentially offer a solid rock foundation on which a telescope could be installed. We note that the practical feasibility of erecting a telescope on a nunatak, or any other rock feature for that matter, depends on a wide variety of factors such as the geometry and slope of the rock surface, the height and size of the feature, the level of glaciation, site accessibility, and the characteristics of the instrument to be installed.

In the southern part of the Ellsworth Mountains is the recently established Union Glacier (UG) logistical hub, where the newly founded Chilean Scientific station and the private tourism and logistics company ALE (Antarctic Logistics and Expeditions) operate during the summer months. The UG camp is reachable by aircraft that land on a blue ice runway located on the glacier itself. Several light aircraft, including Twin Otters, operate from the camp over summer and are capable of deploying instrumentation within a near 1000km radius of the UG hub (the exact range depends on the weight of the payload and other concerns).

Due to its close proximity to the pole, sky brightness in the study area ranges from nautical twilight to astronomical darkness throughout the winter months. Figure 2 shows the estimated incoming solar radiation at the top of the atmosphere for a location at 80°S (close to the UG). The sun begins to dip below the horizon towards the end of May and the onset of continuous twilight or astronomically dark conditions occurs around April 15 and extends until the last days of August. This period is of most interest for astronomy and we have selected the months of May until August (MJJA) as the focus period for this study.

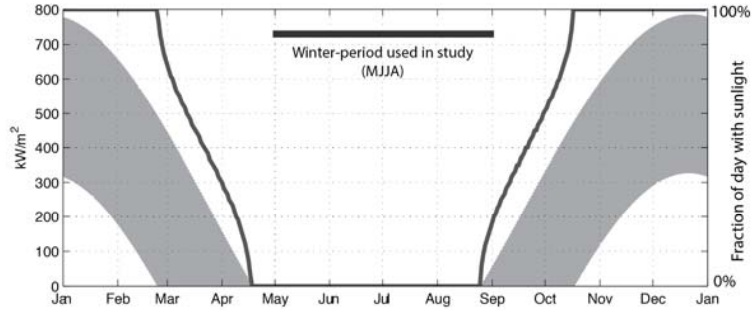

**Figure 2.** Estimated incoming solar radiation at the top of the atmosphere at a point located at 80°S (grey area) along with the fraction of each day in which the sun is above the horizon. The period of near complete darkness from May to August (MJJA) is established as the principal analysis period for this study

## 4. WRF Regional Model

Regional climate models are widely used in both atmospheric research and operational forecasting. Such models provide a complete representation of atmospheric phenomena over a spatial scales ranging from hundreds of meters to thousands of kilometers. RCMs represent the state of the atmosphere on a 3D grid whose lower boundary follows the terrain surface. The grid resolution plays a pivotal role in determining the ability of the model to represent small-scale meteorological features (particularly important in the boundary layer). Higher resolutions are generally better, but the computational cost increases dramatically. For a given spatial region, a model of 1km resolution requires nearly 1000 times more processing power to complete a simulation in the same time as a 10 km resolution run. The computational cost of a simulation also depends on the area of the computational domain and the length of the simulated period. Usually, the configuration of any RCM simulation will involve a tradeoff between these parameters of spatial resolution, domain size and simulation length.

A RCM represents the atmospheric state in terms of a set of 10 or so variables that include wind velocities, temperature, pressure, humidity, cloud water, ice crystals and raindrops. Based on an initial condition (usually derived from large scale gridded metrological analysis) the regional model integrates the equations that govern the time evolution of these variables. During the model integration, the atmospheric state is periodically saved to disk (usually at hourly intervals) for later analysis. On modern computer systems it is feasible to run simulations of the atmosphere for months or even years, and in this way, build up digital climatologies that characterize the spatio-temporal variability of variables of interest.

High resolution simulations have been used in this way in many different contexts, including regional climate change prediction (downscaling, e.g. Murphy,1999) and wind energy prospecting (Probst and Cardenas, 2010). In the analysis of site suitability over Antarctica, regional models have proved decisive in locating the best sites in eastern Antarctica (e.g. Parish and Bromwich, 2007; Swain and Gallee, 2006). To date, most models have been fairly low resolution (30 km and 90km in the case of the previously cited references, respectively). This is justifiable as the topography of most of the continent is rather smooth. However, the topography of WA is more complex and significantly higher resolutions are required if accurate modeling results are to be achieved in this area.

The regional model employed in this study is the polar version of the Weather Research and Forecasting (WRF) regional atmospheric modeling system (Hines and Bromwich, 2008). WRF is a non-hydrostatic weather modeling system appropriate for simulating atmospheric processes over a wide range of spatial scales for any part of the world. The Polar version of WRF was developed by the Polar Meteorology Group at the Ohio State University and includes several optimizations for polar conditions, the most important being a modified surface energy balance model for ice surfaces, and a more detailed treatment of sea ice properties. The polar WRF model, along with its predecessor Polar MMM5, have been widely used in both research and forecast applications over Antarctica (Bromwich et al., 2013) demonstrating good performance in simulating surface winds, humidity and cloudiness.

The specific model configuration used in this study consists of four nested computational domains with spatial resolutions of 27, 9, 3 and 1 km resolution. The 3 km grid covers the entire region of interest for this study and corresponds exactly the spatial region shown in the figure 1. The 1 km resolution model domain (outlined on figure 1b) is centered over the Ellsworth Mountains to the north of the Union Glacier and is intended to better resolve the complex topography of the region. The vertical coordinate of both domains is a stretched terrain following sigma coordinate system, whose lowest level follows the surface topography and uppermost level is fixed at 50 hPa. Model levels are most closely spaced near the surface, the interval between levels being approximately 10 m in the lowest 100 meters. The winter simulation was performed as a single time continuous run over the 5-month period May to September, 2011. Lateral boundary conditions were supplied using 6 hourly 0.5° analyses from the NOAA operational Global Forecast System (GFS). Model data were saved at hourly simulation intervals. Further details regarding the specific options used to configure the WRF simulations are provided in table 2.

The model configuration described above was chosen so as to cover the entire area of interest with sufficient spatial resolution to resolve the key topographic features of the region, especially in and around the Ellsworth Mountains. The tradeoff in this case was the restriction of the simulation period to the single winter season of 2011. Even so, the simulations took about two months to complete on the available computing resources. The relatively short simulation period may introduce uncertainties due to possible year-to-year variability. The only available information regarding inter-year variability is from the surface weather station data (described in section 5), whose data span several winters. We conducted cursory examination of these data and found that at all sites and for all variables the winter-to-winter variability was rather small (< ±10%) and an order of magnitude lower than the spatial variability of these variables within the study area. As such, we believe that inter-annual variability is likely to be a comparatively small source of uncertainty in this study.

The atmospheric data produced by the WRF simulation allow the calculation all of the key diagnostic variables mentioned in section 2. Surface winds are calculated using the wind speed predicted at the first model level, 5.5 meters above the surface. The boundary layer height is continuously diagnosed by the Mellor-Yamada-Janjic scheme (Mellor and Yamada, 1982, Janjić 2002) and these data are included in the saved model output, as is the temperature and humidity at 2m above the surface.

Precipitable water vapor is readily calculated by integrating the 3D model humidity fields in the vertical. Cloud fraction for each grid point is calculated as the fraction of time the integrated cloud water (the sum of all liquid and ice cloud species) in the vertical column that exceeded a small threshold of 0.01 mm (This threshold corresponds to about 10% of the median cloud ice water path over Antarctic latitudes – Feofilov et al, 2015).

## 5. Observational data

### 5.1 Surface observations

Surface meteorological data is sparse over the study area. Standard automatic weather station data (AWS) are available at just two sites maintained by the British Antarctic Survey: The Limbert station at the northwestern edge of the Ronne Ice Shelf and the Sky-Blu airstrip in the base of the Antarctic Peninsula. A welcome source of additional meteorological data in the region stems from the Polar Earth Observing Network (POLENET) geodetic monitoring network (http://polenet.org/). POLENET is made up of precision GPS receivers and seismographs and currently contains over 70 sites. Apart from the geodetic instruments, most POLENET sites have surface meteorological sensors installed. The meteorological data are publically available and add seven additional sites to the available surface data. Both the AWS and POLENET sites measure the temperature, pressure, humidity wind speed and wind direction near to the surface (usually between 1 and 3 m). Table 3 provides summary of the available surface data. Data are often incomplete during winter months.

| WRF Version | 3.3 (Polar WRF) |
| --- | --- |
| Dynamical core | Advanced Research WRF (ARW) |
| Horizontal resolution | 27 km / 9 km / 3 km / 1 km |
| Domain size | 1200 km x 1200 km  / 156 km x 156 km |
| Nesting method | One way |
| Map projection | Polar stereographic |
| Vertical Grid | 42 sigma levels |
| Upper boundary | 50 hPa (approx 20 km) |
| Model topography | RAMP |
| Time step (s) | 40 / 40 / 13.3 / 4.4 |
| Data save interval | 60 minutes on all domains |
| Lateral boundary condition | 6 hour operational analyses from the Global Forecast System |
| Planetary boundary layer | Mellor-Yamada-Janjic |
| Surface layer | Monin-Obukhov |
| Solar radiation | RRTM |
| Long wave radiation | RRTM |
| Cloud microphysics | WRF Single moment (5 species) |
| Convection scheme | Kain-Frisch on 27 and 8 km grids.  No convection on 3 km and 1 km domains |

**Table 2.** Configuration of the WRF model simulations

| Site | Provider | Latitude (°E) | Longitude (°W) | Height (m) | Sampling interval | Data coverage |
|---|---|---|---|---|---|---|
| Sky Blu | BAS | 74.80 | 71.49 | 1554 | 10 min | MJJA 2011 |
| Limbert | BAS | 75.92 | 59.27 | 60 | 10 min | MJJA 2011 |
| Lepley Nunatak* | POLENET | 73.11 | 90.30 | 4 | 30 min | MJJA 2012/2013 |
| Mt Suggs | POLENET | 75.28 | 72.18 | 1107 | 30 min | MJJA 2011 |
| Haag Nunatak* | POLENET | 77.04 | 78.29 | 984 | 30 min | MJJA 2012/2013 |
| Howard Nunatak | POLENET | 86.77 | 77.53 | 1584 | 30 min | MJJA 2011 |
| Wilson Nunatak | POLENET | 80.04 | 80.55 | 663 | 30 min | MJJA 2011 |
| Cordiner Peak | POLENET | 82.86 | 53.20 | 993 | 30 min | MJJA 2011 |
| Whitmore Mountains | POLENET | 82.68 | 104.39 | 2182 | 30 min | MJJA 2011 |

**Table 3**. Weather station data used in this study. A star (*) identifies sites that do not have data for the 2011 model simulation period.

### 5.2 CALIPSO cloud data

The use of satellite derived cloud products is common practice in astronomical site testing but over Antarctica their interpretation complicated for several reasons. Geostationary satellites have low spatial resolution near to the poles and visible images, from which clouds are often most easily discerned, are unavailable over Antarctica during winter. Cloud detection algorithms based on infrared signals often fail to discriminate between cloud top and surface emissions (Serreze and Barry, 2005). The presence of clear sky ice crystal precipitation (diamond dust) is also very difficult to detect from most satellite platforms.

In recent years a new generation of research satellites equipped with active sensors have revolutionized cloud characterization over the polar regions (Grenier et al, 2009, Bromwich et al 2012). An example is the Cloud-Aerosol Lidar and Infrared Pathfinder (CALIPSO) satellite. CALIPSO senses the presence of clouds and aerosols with an active Lidar sensor providing high resolution vertical profiles of cloud occurrence (i.e, cloud height can be determined) that are unaffected by earth's surface and sensitive to optically thin clouds. A limitation of CALIPSO is poor spatial sampling due to its narrow surface footprint. This means that in practice data must be aggregated over long time periods in order to be meaningful (Bromwich et al, 2012). Also, because of the inclination of the satellite orbit, data are unavailable at latitudes higher than 82.5°S.

We obtained a monthly gridded cloud fraction climatology from the GCM Oriented Cloud CALIPSO Product (GOCCP, Chepfer et al, 2010, http://climserv.ipsl.polytechnique.fr/cfmip-obs/). The GOCCP provides gridded global monthly estimates of cloud fraction in 1°x1° grid cells at 250m vertical intervals between the surface and 20km altitude. Cloud fraction is calculated as the fraction of cloud vs. cloud free observations with each grid cell. Maps of total cloud fraction and cloud fraction in lower (0-2000m), middle (2000-6000m) and upper (>6000m) are also provided. To average out noise resulting from the low spatial coverage we make use of all available wintertime data (2007 - 2013).

## 6. Model Results

### 6.1 Winds and boundary layer height

The figure 3a provides an example of observed and simulated wind speeds at Wilson Nunatak, the closest weather station to the Union Glacier. The WRF model does a good job of simulating the observed wind speeds. The strong mean wind speed of 10 m/s is well captured as is the day to day variability (evidenced by a very significant temporal correlation coefficient of 0.62). The observations show considerable variability over short time intervals that are not resolved by the model, presumably due to small scale fluctuations. The prevailing wind direction (not shown) is towards the northwest (204°), roughly in the same direction as the local downslope topographic gradient and is well simulated by WRF (217°).

Figure 4a shows the same variables for the Limbert weather station, located at the northern extreme of the Ronne Ice Shelf. The time series at this location have a very different structure to those at the Wilson Nunatak which is to be expected given the large spatial separation of the two stations. Wind speeds are much lower at the Limbert station with mean values of 5 m/s and few events that exceed 10 m/s. As with the previous station the model does a good job of reproducing the mean value and time variability. Again, the prevailing wind direction of 225° is correctly simulated (model value 243°)

Table 4 presents simple statistics for all stations with data during 2011. Included in the table are mean wind directions calculated from vector average winds. The temporal correlations for wind speed vary between 0.4 and 0.75. These values are in all cases statistically significant (at a 1% significance level). The weakest correlation 0.4 are found at the Howard Nunatak, where a significant difference in mean wind speed and prevailing wind direction are also observed. It seems likely that at this location the winds are strongly influenced by local topographic features not resolved by WRF. Low correlations are also found for the Whitmore Mountains which may partially due to the small amount concurrent observations (22 days) at this site. The correlation coefficients are similar to the values cited in the Polar-WRF forecast model evaluations presented for Antarctica in the work of Bromwich et al, 2013. With the exception of the Howard Nunatak site, prevailing wind directions are well simulated, the model generally capturing the observed direction to within about 30°.

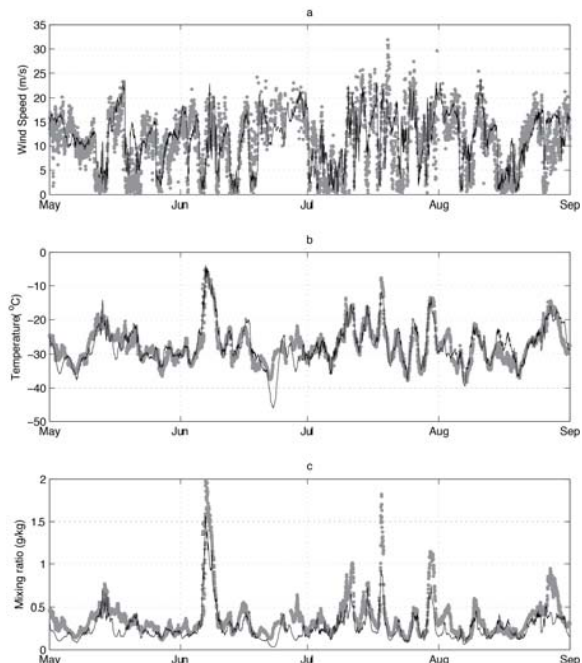

**Figure 3.** Observed (gray dots) and simulated (dark line) wind speed (a, upper panel), temperature (b, middle panel) and specific humidity (c, lower panel) at the Wilson Nunatak POLENET site in the southern extreme of the Ellsworth mountains. The interval between observations is 30 minutes. Model data is at hourly intervals.

For site prospecting, it is the mean value of wind speed that is of most interest. The scatter plot of observed vs. simulated mean MJJA wind speed is presented in figure 5. The plot includes comparisons (2011 for WRF vs 2012-2013) for the two POLENET stations (Haag Nunatak and Lepley Nunatak) that do not have data for 2011. The relationship between observed and simulated wind speed is clearly significant, although biases are present at several sites. The model underestimates the mean wind speed by a substantial margin at the Whitmore Mountains and the Haag Nunatak. This may be in part due to a small sample size in the former and inter-annual variability in the case of the latter.

Maps of the simulated mean wind speed are presented on the figure 6. The large-scale pattern (Figure 6b) is characterized by prevailing winds approximately parallel to the terrain height contours. This is consistent with the near surface flow patterns associated with katabatic winds and the action of the Coriolis force (Parish and Bromwich 1991, 2007). The katabatic flows are most intense on the coastal parts of the northern and eastern flanks of the west Antarctic ridge, where mean winds exceed 10 m/s. At other locations on the continent moderate wind speed of around 6m/s are typical. Wind speeds are lowest over the Ronne Ice Shelf, where the flat terrain presumably limits the development of katabatic flows. The odd looking rectangular features on the edge of the Ronne Ice shelf are artifacts caused at the transition between the ocean and sea-ice region and can be ignored.

Over the Ellsworth mountain region (Figure 6b) a complex local wind field results from the steep topography of the region. A strip of particularly strong winds follows the sharp drop in terrain from the ice plateau to sea level. The strong winds observed at the Wilson Nunatak are associated with this feature. The Union Glacier research station also happens to be located in this particularly windy region. Indeed, the zone of blue ice upon which the UG station runway is located is largely a product of unusually strong winds that result in enhanced ablation over the glacier surface exposing the blue ice regions (Bintanja 1999). Visual inspection of satellite images of the Ellsworth Mountains indicates that the strip of strong wind coincides approximately with areas of blue ice, providing an additional indication that the model wind speed pattern is realistic in the area. Further to the north a second zone of very strong wind speed is associated with the Vinson Massif, which has a height of 3800m on the 1 km model grid. Although these elevated peaks are very windy and so unlikely to prove good candidates for observatory sites, the simulation shows several moderately high regions on the ice plateau immediately to the west where mean winds are much lower. In particular, the zones marked with the symbol * on Figure 6b are all over 1500m high and have mean wind speeds lower than 4 m/s, a value comparable to the conditions found in Mauna Loa (5 m/s; Bely 1987), Paranal in the Atacama Desert (6.5 m/s, Martin et al, 2000) or on the plateaus of Eastern Antarctica (3 m/s; Aristidi et al, 2005b).

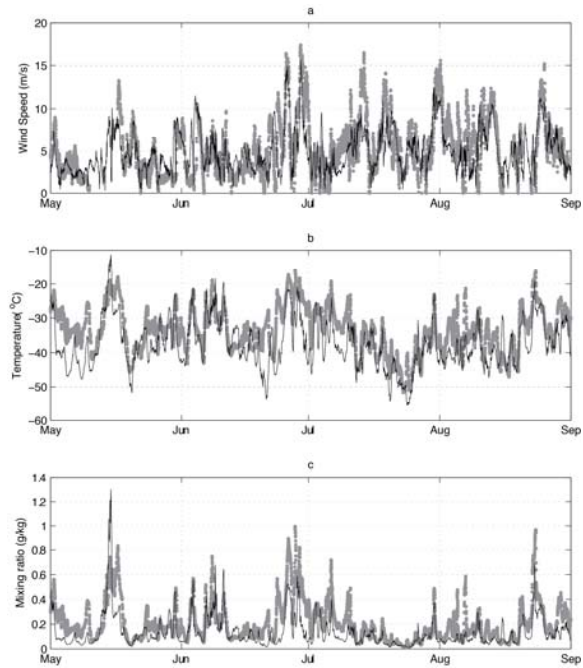

**Figure 4.** Observed (gray dots) and simulated (dark line) wind speed (a, upper panel), temperature (b, middle panel) and specific humidity (c, lower panel) at the Limbert AWS site on the northwestern edge of the Ronne Ice shelf. The interval between observations is 10 minutes. Model data is at hourly intervals.

The MJJA mean planetary boundary layer height (PBL) is shown on figure 7 and exhibits a spatial pattern very similar to that of the mean wind speed. This is consistent that the development of the boundary layer over winter is largely by the generation of mechanical turbulence within the katabatic flows. In the zones of highest winds, the mean PBL height may exceed 500 m. In regions of low wind speeds, the PBL heights are typically between 50 and 100m which is significantly higher than the 20-50m PBL heights that have been measured on Domes A and C (e.g., Agabi et al, 2006). However, we are not to be overly dismayed by this result, as there is considerable uncertainty in the models ability to represent thin nocturnal boundary layer due to the finite model resolution near to the surface and the inherent limitations of its turbulence scheme.

| Site | Wind Speed (m/s) | | | | Wind direction (°) | | Temperature (°C) | | | | Specific humidity (g/kg) | | | |
|---|---|---|---|---|---|---|---|---|---|---|---|---|---|---|
| | N | r | OBS | WRF | OBS | WRF | N | r | OBS | WRF | N | r | OBS | WRF |
| Wilson Nunatak | 103 | 0.62 | 10.1 | 11.0 | 204 | 217 | 122 | 0.85 | -27.4 | -27.4 | 122 | 0.86 | 0.36 | 0.27 |
| Howard Nunatak | 89 | 0.40 | 6.7 | 5.2 | 170 | 95 | 110 | 0.71 | -24.4 | -23.0 | 110 | 0.67 | 0.58 | 0.51 |
| Cordiner Peak | 91 | 0.48 | 7.2 | 7.0 | 85 | 124 | 93 | 0.72 | -26.4 | -31.1 | 93 | 0.60 | 0.34 | 0.2 |
| Whitmore Mountains | 21 | 0.43 | 13.2 | 8.0 | 25 | 1 | 41 | 0.71 | -26.7 | -31.2 | 41 | 0.75 | 0.53 | 0.34 |
| Mt Suggs | 20 | 0.53 | 5.4 | 5.6 | 331 | 301 | 22 | 0.72 | -19.9 | -19.2 | 22 | 0.78 | 0.88 | 0.76 |
| Limbert | 110 | 0.60 | 4.9 | 5.0 | 225 | 243 | 123 | 0.71 | -33.3 | -37.5 | 123 | 0.67 | 0.21 | 0.14 |
| Sky Blu | 16 | 0.75 | 8.0 | 9.8 | 17 | 5 | 16 | 0.78 | -21.6 | -21.1 | 15 | 0.84 | 0.81 | 0.72 |

**Table 4.** Comparison of WRF model with in situ observations. N is the number of days of observations available during the winter (MJJA) of 2011. *r* is the correlation coefficient of all available paired WRF and observed data. OBS and WRF are the mean observed and simulated values of all paired observations.

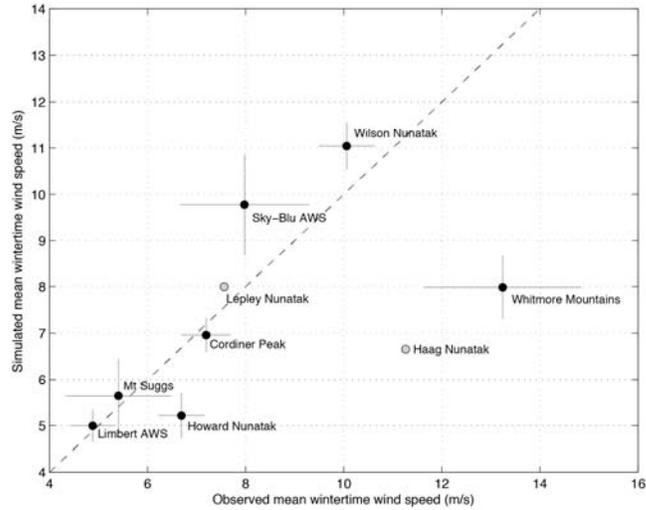

**Figure 5.** Scatter plots of observed vs. simulated scalar average near surface wind speed (black dots) along with estimated 90% confidence intervals (vertical and horizontal bars).  Black dots indicate data points based on the comparison of simultaneous observations during the winter of 2011.  The grey circles indicate points for stations without data during 2011 for which the simulated May to August mean is compared with the long term mean for the same months based data from other years.

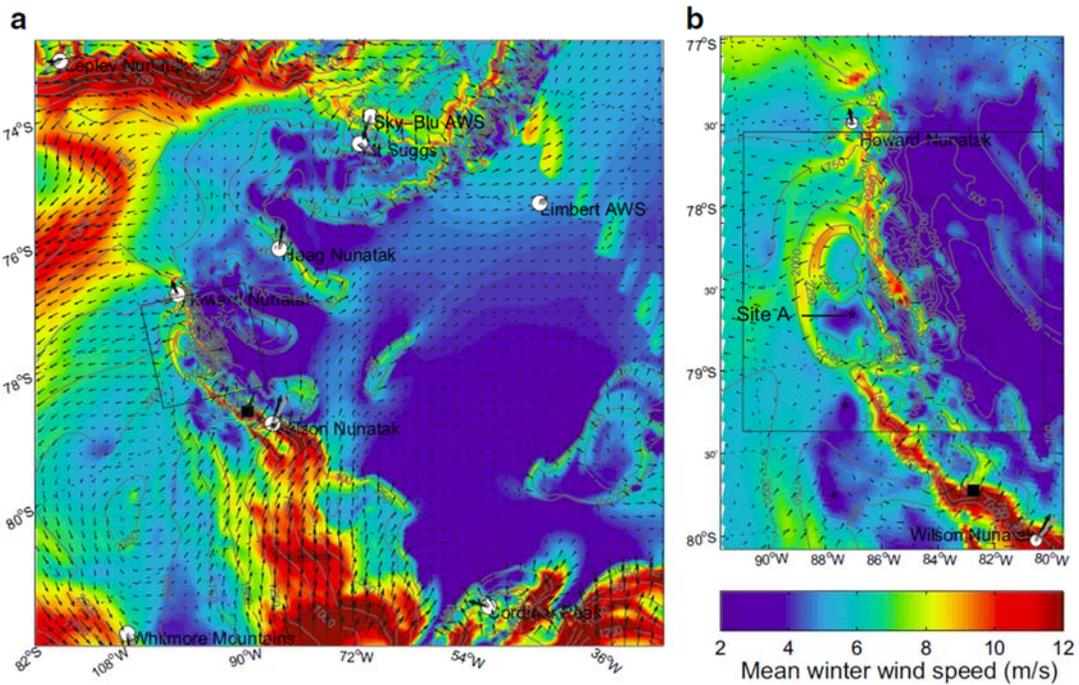

**Figure 6.** Spatial pattern of mean winds for MJJA of 2011. The leftmost panel (a) shows simulated winds over the 3km model domain. Colors represent the mean wind speed and arrows show the vector mean wind directions at 12 km intervals. Vector mean observed (gray) and simulated (black) wind speeds are also shown at the weather station sites.  Here the arrows have been scaled by a factor of 2 for clarity.  The right panel (b) shows a detail over the high spatial resolution domain. The Site A identified in panel b) is the best performing location in the site suitability analysis presented in section 7.

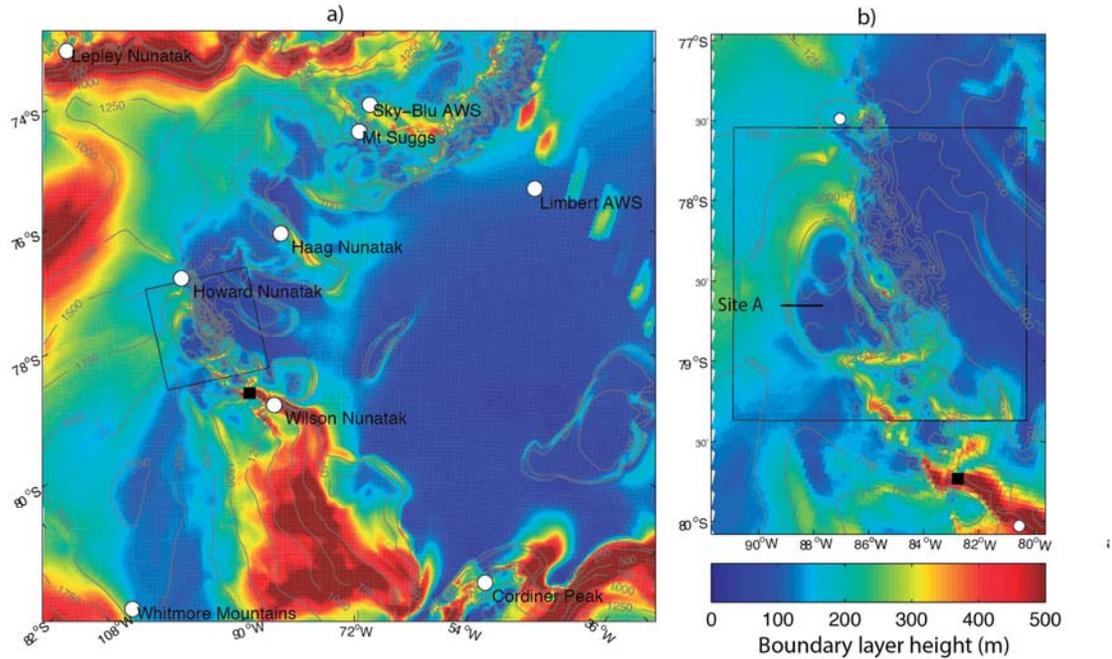

**Figure 7.** Spatial pattern of mean boundary layer height for MJJA of 2011. The leftmost panel (a) shows results over the entire 3km model domain. The right panel (b) shows a detail over the high spatial resolution domain. The Site A identified in panel b) is the best performing location in the site suitability analysis presented in section 7. The black square in both panels is the Union Glacier Hub.

### 6.2 Temperature and water vapor

Examples of observed and simulated temperature and water vapor mixing ratio at Wilson Nunatak are shown in figures 3b and 3c. The model shows a good performance for both variables, with temporal correlations of 0.85 and 0.86 respectively, significantly higher than the correlation obtained for wind velocity. The model appears to be have bias towards under prediction of the mixing ratio (i.e., the model is too dry). It is of interest to note the relatively constant nature of the temperature data during the winter months, with no clear evidence of a mid-winter minimum. This is an expression of the so called "coreless" winter in west Antarctica, and has been attributed a southward transport of heat and moisture by small scale cyclonic systems, that balances radiative cooling at the surface (Connolley and Cattle, 1994). The intrusion of warmer, moist weather systems is also suggested by the periods of intense warming and moistening that occur several times during the winter period. The most dramatic is an episode centered on June 6 during which the temperature rose to close to 0°C and the mixing ratio to 1.5 g/kg, nearly 5 times its mean winter value. Further study is required to determine the origin of these episodes but it seems probable that they are associated with intrusions of marine air masses (Nicolas and Bromwich, 2011) and given the high humidity involved are likely be associated with cloud cover and poor astronomical observing conditions.

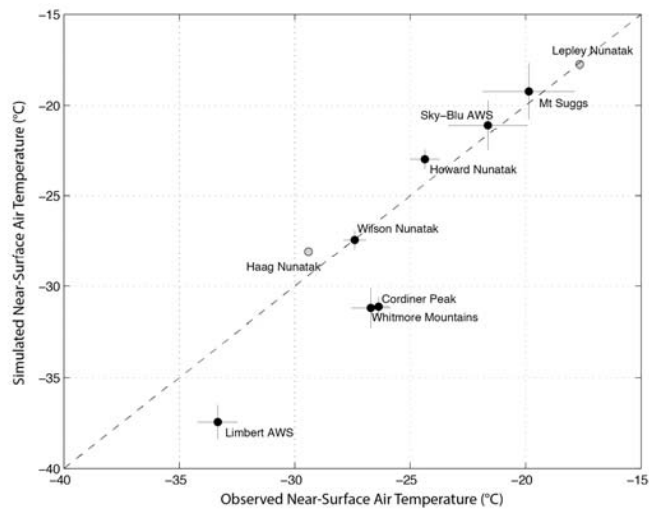

**Figure 8**. Scatter plot of observed vs. simulated mean near surface temperature. The meanings of the symbols are the same as for figure 5.

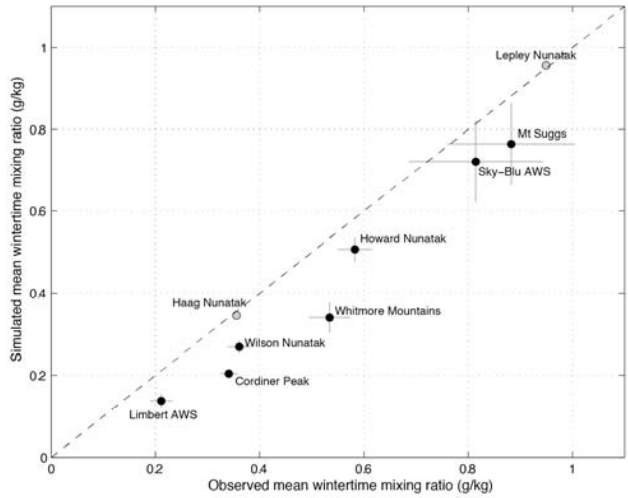

**Figure 9**. Scatter plot of observed vs. simulated mean near surface specific humidity. The meanings of the symbols are the same as for figure 5.

The temperature at the Limbert site (Figure 4b) shows much lower values than the Wilson Nunatak and a more episodic time variation. Periods of unusually high temperature and humidity occur at this site but a less pronounced compared to the Wilson Nunatak site. The model has a clear temperature bias at this station, the mean simulated value being -37.5°C compared to the observed value -33.3°C. The same may be said of the mixing ratio whose mean value is underestimated by the WRF simulations.

Temperature and humidity statistics for all sites are displayed in Table 4. Temporal correlations of around 0.7 to 0.85 are typical for both variables. Scatter plots of observed vs. modeled mean temperature and mixing ratio are presented in figures 8 and 9, respectively. The plots demonstrate a clear linear relationship, especially for mixing ratio, which also shows a systematic low bias of about 0.1 g/kg. The temperature results are generally very good although the model exhibits a strong negative bias of about 5° at 3 sites (Limbert, Whitmore mountains and Cordiner Peak).

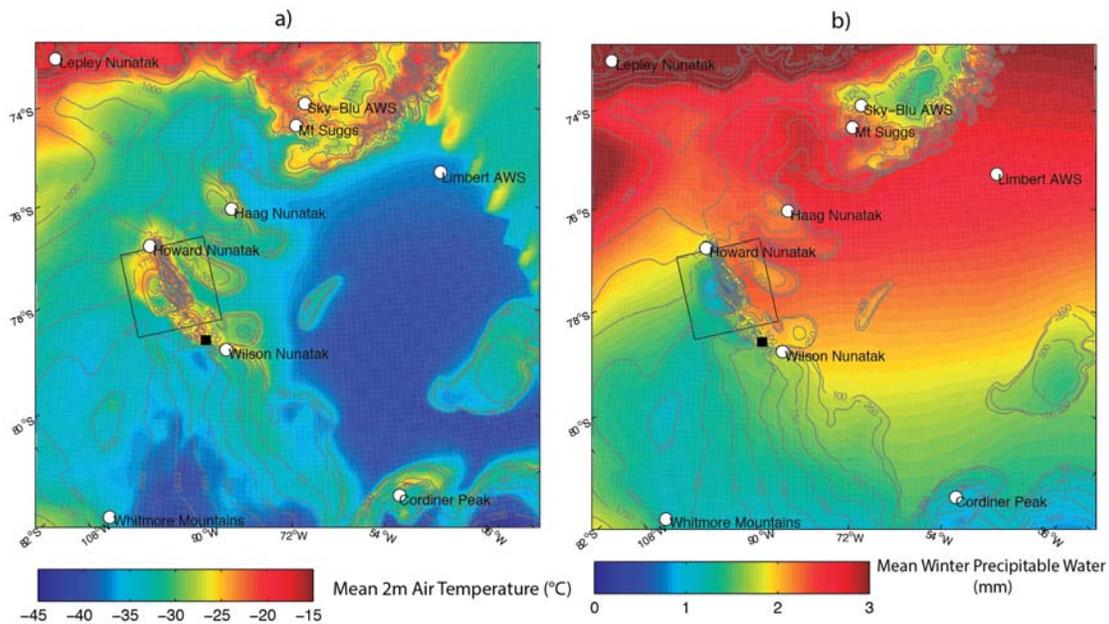

**Figure 10.** Spatial pattern of mean air temperature (a) and precipitable water (b) for MJJA of 2011. The leftmost panel (a) shows results over the entire 3km model domain.

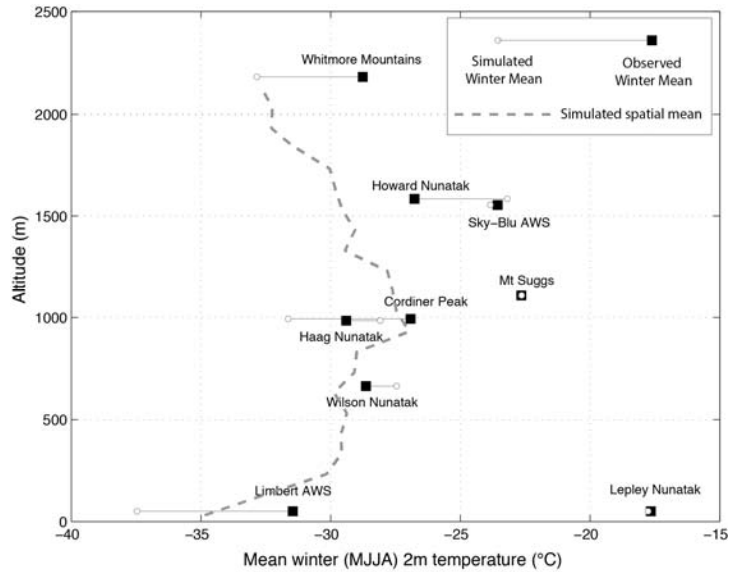

**Figure 11.** Variation of mean MJJA 2m temperature with height at weather station sites within the study area. Black squares and white circles connected by lines indicate observed and simulated temperatures, respectively. The gray dashed line shows the spatial mean surface temperature as a function of height based on the results of the 3 km WRF simulations.

The simulated spatial patterns of temperature and precipitable water (PW) are shown in figure 10. Perhaps the most notable feature of the temperature map (Figure 10a) is pool of very cold air over the Ronne Ice Shelf that contrasts with the warmer temperatures over the higher altitude surrounding topography. This feature points to the presence of a deep temperature inversion over the ice shelf during the winter months. The observed profile of MJJA temperature against terrain height for measured and modeled data is shown in figure 11. The data indicate that the inversion height is at around 1500m. This estimate is to some degree corroborated by inspection of the simulated temperature pattern whose spatial mean generally shows maximum values between 900 and 1200 m (see dashed line in figure 11). The milder temperature at the Lepley Nunatak, at sea-level on the northern coast of Ellsworth Land confirms that the inversion is confined to the eastern side of the Antarctic Peninsula. The distinctive temperature pattern over the Ronne Ice Shelf and Weddell Sea has been documented several past works (Schwerdtfeger 1975; Morris and Vaughn 2003). Schwerdtfeger examined radiosonde data from early expeditions (around 1924) in the Weddell sea slightly to the north of our area and estimated inversion height of around 800m. Morris and Vaughan (2003) examined surface met records and subsurface snow temperatures and obtained a spatial pattern qualitatively similar to the WRF result. The temperature inversion likely occurs as a result of sustained radiative cooling over the ice-shelf during the winter months while the anomalously warm regions in the mountains may be related to Foehn effects (e.g., Steinhoff et al, 2013) or subsidence warming due to katabatic outflow (Simmonds and Law, 1995).

The precipitable water (PW) shows a much simpler spatial pattern (Figure 10b), characterized by a north to south gradient and strong decrease with terrain elevation. In the northern part of the study PW at sea level is between 2 and 3 mm, comparable to the annual mean values observed at observatories in Chiles Atacama Desert. Within the Ellsworth Mountains mean PW is lower than 1 mm for sites above 1500 m altitude.

### 6.3 Cloud Fraction

Figure 12a shows the MJJA cloud fraction derived from the WRF simulations. The model predicts a significant west to east gradient in cloud cover with the cloudiest conditions (70%) found to the west and clearest conditions to the south east (where cloud fraction diminishes to under 30%). A considerable increase in cloud frequency is predicted over the Ellsworth Mountains. These spatial patterns are quite similar to those derived by Nicolas and Bromwich, 2011 from the Polar MM5 (predecessor to WRF) AMPS forecast system.

The figure 12b shows the estimated total cloud frequency from the CALIPSO satellite. While the CALIPSO product also shows an east west gradient in cloud frequency across the study area, the mean cloud frequency is much higher and the gradient weaker. According to the CALIPSO product, cloudiness varies from over 80% in the west to around 50% in the far east. The discrepancy could in part be explained by the different time periods being compared (the CALIPSO climatology covers the MJJA for 2006-2013 while WRF is for 2011 only), but it seems unlikely that the very large differences can be entirely put down to inter annual variability. Bromwich et al, (2013) noted that the Polar WRF model tends to underestimate cloud occurrence and it appears that the same bias has affected our simulations also.

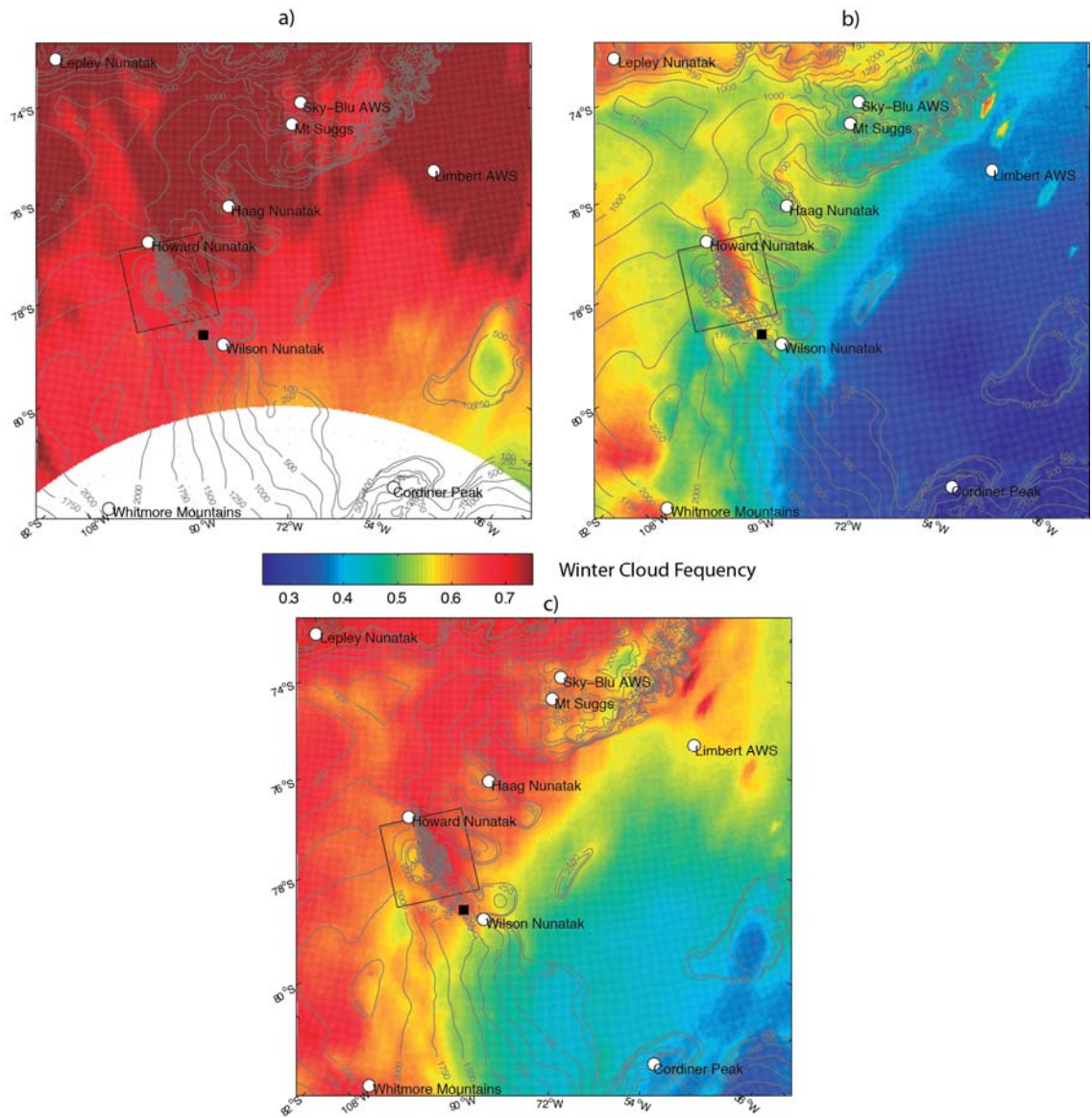

**Figure 12**. Simulated and CALIPSO spatial pattern of wintertime (MJJA) mean cloud fraction over the study area. The WRF model data (upper right panel) are based on the 3km and 1 km simulations for the year 2011. The CALIPSO data (upper left panel) have a spatial resolution of 1° and are based on 8 years of data from 2006 until 2013. The CALIPSO data is the total cloud fraction considering cloud anywhere from the surface to 20 km altitude. The lower panel shows the mean of the WRF and CALIPSO cloud fraction that is used in the site suitability analysis presented in section 7.

The very high cloud fractions suggested by CALIPSO, which we consider to be a very reliable source of cloud data for Antarctica, has significant implications regarding the overall on suitability of this part of WA and it is of interest to place the result in a larger scale context. The figure 13a shows CALIPSO total cloud fraction over Antarctica and also the southern cone of South America. The maps include the locations of two existing major observatory sites in Chiles northern desert (Cerro Paranal and Cerro Pachon) along with Dome A site: a current hotspot for Antarctic astronomy. The cloud fraction data show pronounced minima over both dome A and northern Chile, where the frequencies of around 30% in both cases are about half that observed in western Antarctica. Figure 13b shows the frequency of high cloud occurrence, defined as clouds detected above 6000m. Here the contrast between western and eastern Antarctica is particularly dramatic, Dome A showing a pronounced minimum in high cloud frequency and WA Antarctica a clear maximum. The WA maximum is actually centered to the west of the West Antarctic ice divide but extends sufficiently eastwards to adversely affect our study area. The large scale cloud pattern has been discussed in Bromwich et al, 2012 who attributed the pattern to synoptically driven vertical motion over the WA's topography. The figure 13c shows a vertical cross section of mean cloud fraction between the dashed lines shown in figures 13a and 13b. The profile shows that the cloud frequency is greatest at upper levels over the study area with cloud occurrence most frequent at about 9 km altitude. At the Dome A and Chilean sites, clouds are rare above the observatory altitudes. The Chilean sites demonstrate clearly how the location of the observatories on mountaintops allows them to avoid the impact of frequent low level marine clouds below 2000m.

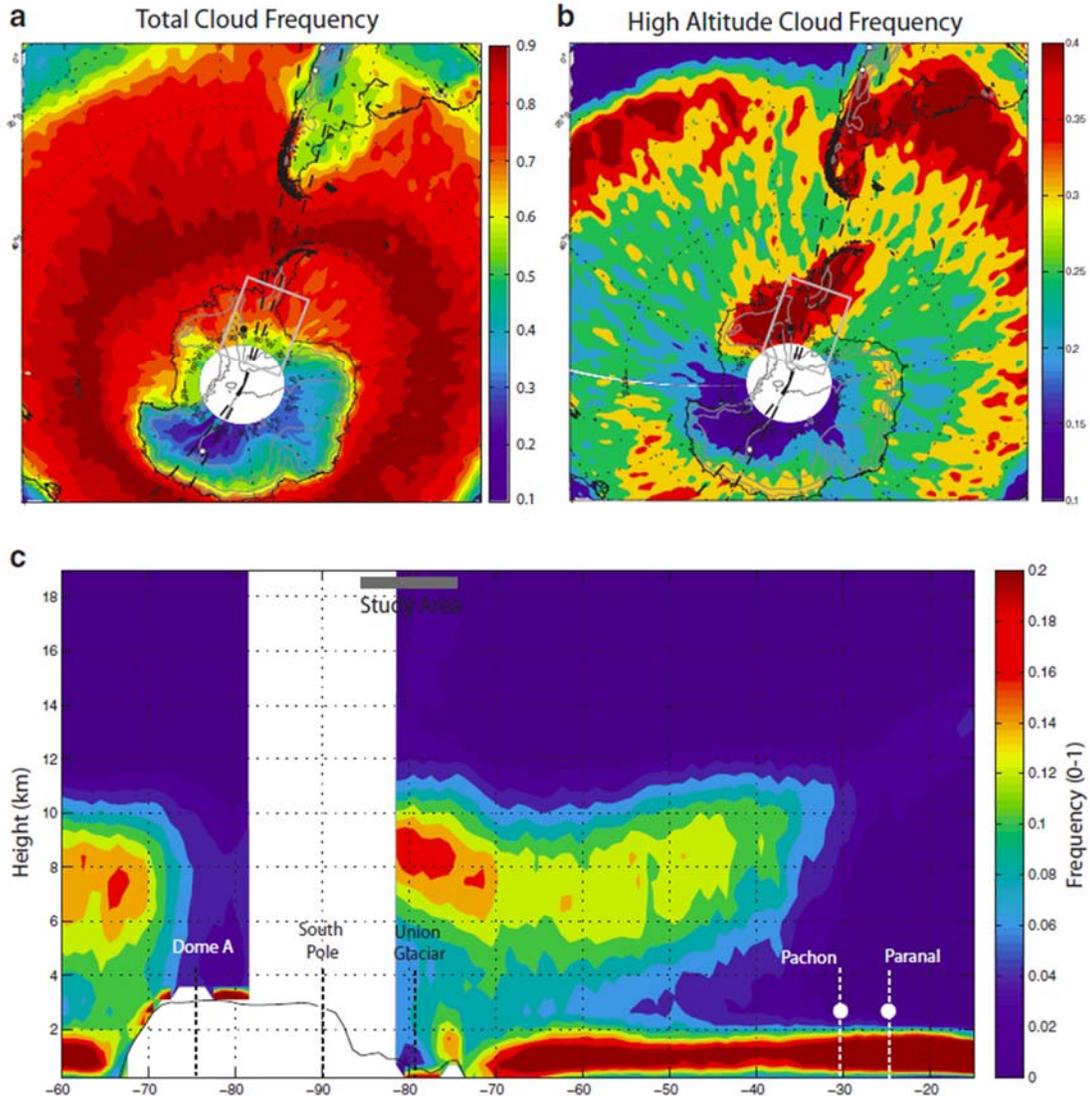

**Figure 13**. Large scale MJJA cloudiness patterns from CALIPSO from 2006-2013. The upper left panel (a) shows total cloud cover over Antarctica and the southern cone of south America. Black and white dots show several existing observatory sites including Paranal and Pachon, two of Chiles most important observatory sites and Dome A in Eastern Antarctica. The gray squares show the study area in western Antarctica. The dashed lines indicate the location of the cross section shown figure 13c. The upper right figure (b) shows the mean high cloud occurrence fraction, where high clouds are defined as those that occur between 6 and 20 km. The lower panel shows a vertical cross section of cloud occurrence frequency averaged between the dashed lines shown in panels a) and b). The horizontal axis of this plot shows latitude; starting from (60ºS, 125ºE) and extending to (15ºS, 70ºW) (just north of Chile). The same points of interest and their respective heights above sea level are marked

## 7. Site suitability model

Thus far we have examined the relevant meteorological variables independently. However, it is clear that the spatial patterns these variables are different and the best sites for astronomy will be those that bring together the most favorable set of attributes. How do we combine the model results in order to objectively reveal the "best" locations? To answer this question, we develop a simple site suitability model that assigns an objectively defined score to each geographical location within the study area. A qualitatively similar methodology was applied by Schöck et al (2011) to evaluate the merit of candidate sites for the Thirty Meter Telescope project.

In a general sense, a site suitability model is a tool for identifying locations in a landscape where multiple criteria overlap in geographic space. The use of site suitability models (hereafter SSM's) is commonplace for users of geographic information systems (GIS) and provides a convenient way of ranking sites according to the degree to which they meet a given set of criteria. An overview of the many different SSM approaches may be found in Malczewski, 2004. The approach used in this study falls into simple class of SSM models based on map algebra (Tomlin 1994). It works by assigning a site suitability score (S) that ranges between 0 (unsuitable) and 1.0 (optimal). S is calculated as the product ($\Pi$) of a set of weights (also in the range 0 and 1) that are determined for each factor Xi by applying a transfer function fi(Xi). That is:

$$S = \Pi_i f_i(X_i)$$

The transfer function is defined as a simple piecewise curve that varies linearly between prescribed limits. For example, the transfer function for surface winds (x) is defined as:

$$f(x) = \begin{cases} 1, & x \leq 4 \\ \dfrac{8 - x}{4}, & 4 < x < 8 \\ 0, & x \geq 8 \end{cases}$$

This equation states that any site where the mean winter wind speed exceeds 8 m/s is considered unsuitable, while point with a mean wind speed of 4 or less is considered to be excellent for astronomy and is assigned a perfect score. The score varies linearly between these limits. A map of f(x) for wind speed along with the associated weight function map is shown in figure 14a. Note that the operation to combine multiple weights is multiplicative, so that if any particular factor is evaluated to be zero, the final score will also be zero.

The table 5 summarizes the factors that were considered in the analysis along with the upper and lower limits used to define the weighing functions f(X) in each case. The list includes both meteorological factors along with other "logistical" aspects that we expect to play a pivotal role in determining the real world site suitability in the study area.

Meteorological factors that were considered include the mean wind speed (as described previously), the boundary layer height, integrated water vapor, surface temperature and cloud cover. Limits for the lowest and highest scoring values were chosen based on the simulated spatial variability and correspond roughly to the 20% and 80% percentiles of the distribution on values within the study area. The water vapor and surface temperature are considered to be non-essential factors, in the sense that it could still make sense to make astronomical observations at the site even if the mean values of these factors are far from optimal, and are assigned a lowest possible score of 0.5 instead of 0.0.

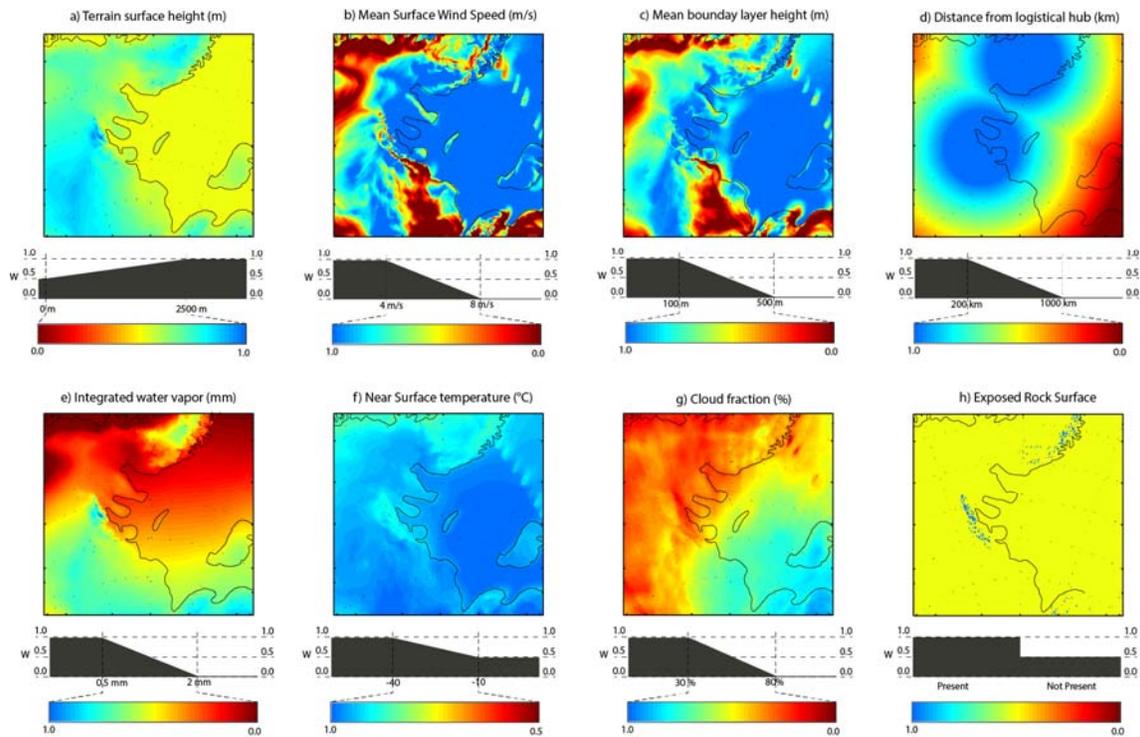

**14.** Individual weighting maps for all of the factors considered in the suite suitability model. Each map shows the evaluated piecewise linear transfer function whose values range from 0.0 (white) to 1.0 (dark gray). The corresponding transfer function is shown below each map. The same transfer function parameters are also provided in figure 4.

| Factor | Worst | | Best | | Type | Value at "best" site | |
|---|---|---|---|---|---|---|---|
| | Value | Score | Value | Score | | Site B Meteorological factors only | Site A Full set of factors |
| Surface wind speed | > 10 m/s | 0.0 | < 5 m/s | 1.0 | Meteorological (E) | 4.2 m/s (1.0) | 4.7 m/s (1.0) |
| Boundary layer height | > 500 m | 0.0 | < 100 m | 1.0 | Meteorological (E) | 42 m (1.0) | 74 m (1.0) |
| Cloud fraction | > 80% | 0.0 | < 30% | 1.0 | Meteorological (E) | 38% (0.84) | 60% (0.4) |
| Integrated water vapor | > 3 mm | 0.5 | < 0.5 mm | 1.0 | Meteorological (N) | 0.8 mm (0.88) | 1.1 mm (0.72) |
| Surface temperature | > 0 °C | 0.5 | < -40°C | 1.0 | Meteorological (N) | -32°C (0.9) | -30°C (0.87) |
| Terrain height | 0 m | 0.0 | > 3000 m | 1.0 | Meteorological (N) | 1720 m (0.79) | 2200 m (0.86) |
| Rock surface | Not present | 0.5 | Present | 1.0 | Logistical (N) | Not present | Present (1.0) |
| Distance from operational hub | > 800 km | 0.5 | < 200 km | 1.0 | Logistical (N) | 678 km | 148 km (1.0) |
| | | | | | | Global Score | |
| | | | | | | 0.52 | 0.22 |

**Table 5.** Factors and criteria used in the site suitability model. The model translates the physical value of each factor to a score within the range [0,1]. The criteria for the lowest scores are given in the column labeled "Worst". The lowest possible score is 0.0 in most cases. Variables that are considered *non-essential* have weighting functions with lower bounds of 0.5. The criteria for the highest possible score (always 1.0) is given in the column labeled "Best". The weighting function varies linearly between these limits. All meteorological variables correspond to mean wintertime values (may to august). The last two columns show factor values and associated scores at the best scoring sites taking into account meteorological factors only (Site B in extreme south-east of the study area) and all both meteorological and logistical factors (Site A situated at a nunatak west of the Ellsworth mountains) The overall scores for these sites are shown in the final row of the table

In the case of the boundary height and cloud cover variables some additional comments are required: The boundary layer height (BLH) has been included in the analysis for two reasons: Firstly, low BLH indicates that surface related turbulence is confined to a thinner layer. Although turbulence in a thin boundary layer may still lead to poor seeing conditions it is easier to correct using modern ground-layer adaptive optics systems (e.g Travouillon et al, 2009). Secondly, if the BLH is lower it becomes increasingly feasible to construct telescopes that are above this layer. Thus lower BLH positively influences site suitability. The use of the 10% and 90% percentiles to define the scoring range leads to lower and upper limits of 100m and 500m respectively. The values are rather generous if interpreted in an absolute sense, as a PBL height of even 100 m is likely to pose great difficulties to practical telescope installation and performance. However, given the well documented difficulties of simulating the nocturnal boundary layer with mesoscale models (e.g. Holtslag et al (2013), Steeneveld (2014)), combined with the lack of in-situ measurements to evaluate the performance of the simulated BLH, we believe it is reasonable to make use of these thresholds as a means of distinguishing areas of lower and higher BLH.

The cloud cover estimates derived from the WRF model tend to underestimate the mean cloudiness compared to the CALIPSO satellite product. On the other hand, the low resolution CALIPSO data lacks the richness of detail of the WRF product (which includes several realistic looking small scale features) and has no data in the southernmost sector of the study area. In an attempt to combine the favorable aspects of both datasets we defined a combined cloud product as the simple average of the WRF and CALIPSO mean cloud fraction fields, extrapolating the southernmost CALIPSO data southwards to fill in the missing data region. The resulting mean cloud cover (figure 14c) ranges from over 70% in the coastal region in the northeast to values below 35% in the southeast near Cordiner peak.

The non-meteorological factors in table 5 include the terrain height, surface type (ice or rock) and the distance from operational scientific bases (specifically the Union Glacier and Sky Blu aerodrome on the peninsula). The topography height is included for the simple reason that, in general, all atmospheric impacts on the astronomical data will diminish with altitude. As such, all other aspects being equal, it is almost always desirable to place observatories at high altitudes if possible. The factor labeled "distance from hub" assigns each location a score that depends on the minimum distance to either of the two operational hubs (Union Glacier and Sky Blu airstrip) within the study area. The "surface type" factor has a different functional form to the other factors and is defined as 1.0 for model grid points where exposed rock is present, and 0.5 where only ice surfaces are available. Note that all the logistical factors are considered to be non-essential and have minimum values of 0.5.

The transfer functions were evaluated at each WRF model grid point in the study area and the individual f(X) maps are presented in the figure 14. As expected, the weighting patterns are similar for surface winds and boundary layer height but are quite different for all other factors. The map of the suitability score obtained by including just the meteorological factors (wind speed, boundary layer height, temperature, humidity, terrain elevation and cloud fraction), is shown in the figure 15. The map suggests that the most promising sites for astronomy are found in 5 broad geographical areas: The ice plateau known as Palmer Land at the base of the Antarctic Peninsula, the upper part of the western flank of the west Antarctic ice divide, the southeastern sector of the Ronne-Fincher Ice Shelf, the Pensacola Mountains in the south eastern part of the study area, and the ice plateau to the west of the Ellsworth mountains. The highest scoring point (labeled "Site B"), whose value of 0.52 is only half the optimum score, is located the Pensacola Mountains in the extreme south east of the simulation domain at 1720 m altitude. Table 5 shows the results for individual variables at site B, which indicate that the site combines low values for winds, boundary layer height and humidity. Its key advantage however is its relatively low value for mean cloud fraction, which, at 38%, is amongst the lowest of any point within the study area. Note that the CALIPSO cloud data do not extend as far south as the Pensacola Mountains so this cloudiness estimate is largely based on the WRF model simulations.

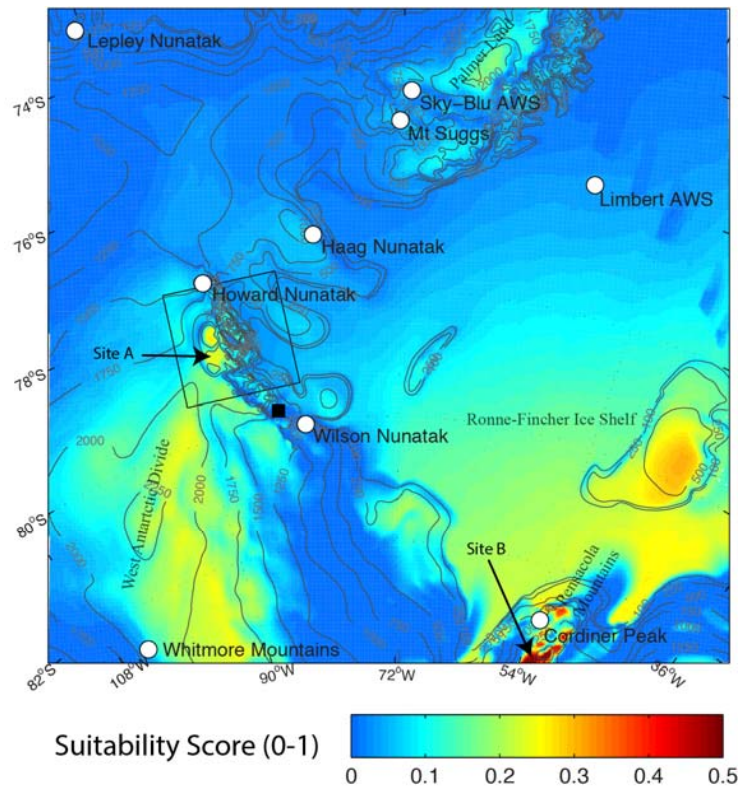

**Figure 15.** Site suitability score map derived from the SSM including only meteorological factors and terrain elevation. Scores range from 0.0 to 0.52. The black square is the Union Glacier hub. The arrows points to highest scoring sites, Site A and Site B, located to the North of the Union Glacier hub, and in the southeastern sector of the study area, respectively.

The map of the overall site suitability score S, obtained by multiplying together each of the maps in figure 14 including the logistical factors of rock surfaces and distance to operational hubs, is presented in the Figure 16. Many of the highest scoring areas mentioned previously now receive a much lower score due to their great distance from the operational hubs. For example, the entire Pensacola range is effectively discounted due to the > 600 km distance from the Union Glacier (nearest operational hub). The highest scoring locations in this case are found on the ice plateau to the west of the Ellsworth Mountains and are associated with Nunataks that protrude from the high altitude ice surfaces. The last column on Table 5 shows the results for the best scoring point, "Site A", shown on Figure 1b. The sites is high scoring partly because of the logistical factors considered but also has favorable conditions in terms of relatively low winds speeds (4.7 m/s) and boundary layer heights, low moisture and high altitude. It must be recognized that the overall score of 0.22 is rather low compared to maximum possible values of 1.0, mainly due to the high levels of cloudiness (~60%) over the region surrounding the Union Glacier hub.

The results of the analysis presented in this section are based on weighting factors have been defined in a more or less ad hoc manner and readers could well question the appropriateness of particular threshold values that were used. We experimented with making small adjustments to the criteria and found the results to be quite robust in the sense that our overall conclusions were not significantly altered as a result of the changes. However, it must be recognized that there is no objective way to determine the most appropriate thresholds to use. We can only stress that the purpose of the exercise is not to obtain a rigorous estimate of the "real" site suitability (if such a thing exists) but is simply to provide a succinct way of visualizing the most promising locations within a given geographical region according to a reasonable set of rules. The aim of the analysis is to provide guidance and definitive results can only be obtained from actual field data.

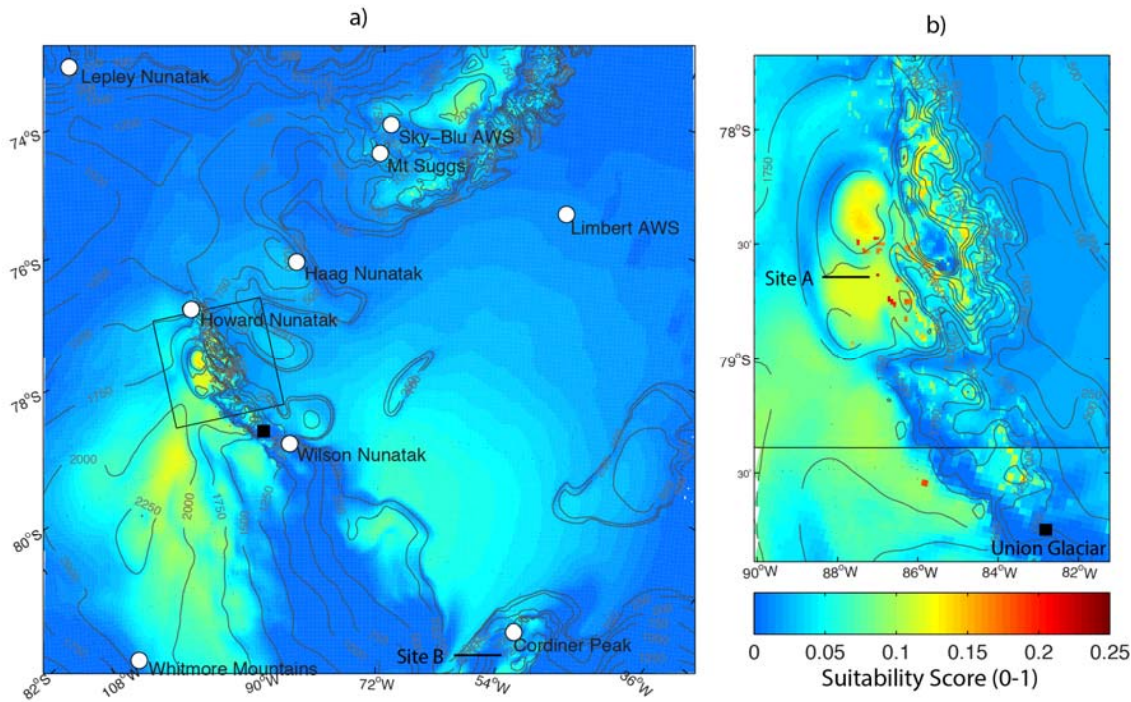

**Figure 16.** Final site suitability score map derived from the SSM including both meteorological and logistical factors. Scores range from 0.0 to 0.25. The best sites are highlighted with dashed circles, are associated with nunataks located on the ice plateau to the west of the Ellsworth Mountains. These sites combine favorable meteorological conditions, the presence of rock surfaces, high terrain and a close proximity to the Union Glacier logistical hub. The location of the best scoring site is marked as "site A".

## 8. Summary

In this paper we have presented a novel application of the Weather Research and Forecasting model to the problem of astronomical observing site suitability assessment in West Antarctica. To our knowledge this is one of the first times that a regional model has been expressly applied to this purpose and also the first time that high resolution simulations have been conducted over the Ellsworth Mountains and its surroundings.

The work presented here is based upon nested WRF simulations at 3 km and 1 km resolution from May to August of 2011. The model outputs were evaluated against data from nine automatic weather stations and CALIPSO data products, with generally favorable results. A simple scoring scheme was applied to obtain a map of site suitability. The scoring system takes into account not only the meteorological data produced by the WRF model but logistical factors that are also of crucial importance when looking for real world sites

The most important findings may be summarized as follows:

• The WRF model performs well in predicting the spatial and temporal variability of wind speed and direction. Temporal correlations significant and between 0.4 and 0.65, and the mean simulated wind speed is typically within 1-2 m/s of the observed mean.

• WRF temperatures are generally well simulated although three sites show large negative biases (~ -5°). Both model and observed data confirm the presence of a deep (1000 - 1500m) temperature inversion over the Ronne Ice Shelf during the winter season. Highest temperatures occur on the coast on the western side of WA ice divide and where the topography intersects the top of the inversion.

• WRF mean surface mixing ratio shows an excellent spatial correlation but has a negative bias, the WRF model being drier than observed.

• The WRF simulations of cloud occurrence show realistic looking spatial features including a local maximum over the Ellsworth Mountains and decreasing cloud cover from west to east over the study area. However, comparison with CALIPSO climatology for the period 2006-2013 suggests that the WRF cloud frequencies are significantly underestimated.

• The CALIPSO satellite data show that Western Antarctica is much cloudier than the Eastern Antarctica. The increased cloud frequency is principally due to a pronounced anomaly in the frequency of high clouds. Overall, cloud occurrence frequency in the study area is nearly twice that of dome A in Eastern Antarctica. The presence of frequent high clouds places significant doubts on the possibility of doing quality optical and near-IR astronomy in this part of WA as there is no way to avoid these clouds by going to higher ground elevations.

• The site suitability model proved to be useful means of combining the meteorological data from the WRF model and other non-meteorological factors. The results of the analysis allowed us to detect the most promising site locations in a quantitative and reproducible framework.

The model results in this study have laid the foundations for future astronomical work in western Antarctica, where the presence of the recently inaugurated science base in the Union Glacier will doubtless open the door to a host of new science projects in upcoming years.

An interesting option to consider is the possibility of astronomical project in the immediate vicinity of the Union Glacier base. Unfortunately, our results predict strong winds over the Union Glacier site and its immediate surroundings, a feature confirmed by the observations at the nearby Wilson Nunatak. These winds are associated with strong boundary layer turbulence and for this reason alone we can conclude that Union Glacier science base is unlikely to be an appropriate site for astronomical work and we must look further afield for viable sites in the region. Another area that a priori was considered to be of potential interest was the highest peaks of the Ellsworth range, including the Vinson Massif itself. Once again, the model rules out this possibility due to its prediction of strong winds and enhanced cloudiness over the mountains. (In fact, logistical considerations alone are probably sufficient to preclude the possibility on Mt Vinson or neighboring peaks).

Based on the results of the site suitability model we instead arrive at the unexpected conclusion that the most promising sites are actually located at the somewhat innocuous Nunataks that poke out of the high ice plateaus to the west of the Ellsworth mountains. The highest scoring Nunatak shows favorable atmospheric conditions: low moderate wind speeds, low moderate boundary layer and water vapor, moderate temperatures at an effective altitude that is comparable to many large observatories in the subtropics. It also appears feasible to reach the site by plane and anchor instruments to the rock surface of the Nunatak: a unique advantage of the WA sites. Unfortunately, site performance is let down by the high cloud occurrence fraction of around 60%. This is a disadvantage that appears ubiquitous over western Antarctica. Indeed, even if we relax our logistical constraints and base site suitability on only the meteorological factors, the best site, found in the Pensacola Range some 650 km to the east of the UG hub, still has a rather high mean cloud fraction of 38%.

Future work must involve the collection of field data at the detected promising sites to confirm or deny model results. A potential field campaign would ideally include, apart from standard meteorological sensors, instruments to measure cloud cover, turbulence and seeing. The experiences described in Steinbring et al (2010) in the Canadian Arctic provide an interesting example as to how such a field campaign could be carried out. Given that there exist doubts as to the viability of the sites, the most prudent course of action may be to collaborate with other groups also interested in obtaining meteorological observations in the area (glaciologists, geodesists and climatologists being obvious candidates). We close by emphasizing that the modeling techniques described in this work are in principal applicable in any part of the world and are becoming increasing popular within the astronomical community, both for site selection and operational forecasting (e.g. Masciadri and Lascaux, 2012, Giordano et al, 2014). We trust that our example will help further the establishment of regional climate models as a standard tool in astronomical site selection process.

**Acknowledgements:** Patricio Rojo and Mark Falvey acknowledge funding by the Instituto Antártico Chileno project INACH G19_11. We also acknowledge support from the BASAL CATA Center for Astrophysics and Associated Technologies PFB-06.